\documentclass[conference]{IEEEtran}
\IEEEoverridecommandlockouts
\usepackage{cite}
\usepackage{amsmath,amssymb,amsfonts}
\usepackage{algorithmic}
\usepackage{graphicx}
\graphicspath{{images/}}
\usepackage{textcomp}
\usepackage{subcaption}
\usepackage{float}
\usepackage{xcolor}
\def\BibTeX{{\rm B\kern-.05em{\sc i\kern-.025em b}\kern-.08em
    T\kern-.1667em\lower.7ex\hbox{E}\kern-.125emX}}
\begin{document}

\title{Enhancing Ultrasound Molecular Imaging: Toward Real-Time RPCA-Based Filtering to Differentiate Bound and Free Microbubbles\\
\thanks{This work was supported by the Stanford Cancer Imaging Training (SCIT) program, funded by the National Cancer Institute (NCI) through training grant T32-CA009695, and the National Institute of Biomedical Imaging and Bioengineering (NIBIB) under grant R01-EB031799.}
}

\author{\IEEEauthorblockN{Hoda S. Hashemi, Dongwoon Hyun, Nathan Nguyen, Jihye Baek, Arutselvan Natarajan, Farbod Tabesh,}
\IEEEauthorblockN{Andrew Andrzejek, Ramasamy Paulmurugan, and Jeremy J. Dahl}
\vspace{3mm}
\IEEEauthorblockA{\textit{Department of Radiology, Stanford University, Stanford, CA, USA}}
}

\maketitle

\begin{abstract}
Ultrasound molecular imaging (UMI) is an advanced imaging modality that shows promise in detecting cancer at early stages. It uses microbubbles as contrast agents, which are functionalized to bind to cancer biomarkers overexpressed on endothelial cells. A major challenge in UMI is isolating bound microbubble signal, which represents the molecular imaging signal, from that of free-floating microbubbles, which is considered background noise. 
In this work, we propose a fast GPU-based method using robust principal component analysis (RPCA) to distinguish bound microbubbles from free-floating ones. We explore the method using simulations and measure the accuracy using the Dice coefficient and RMS error as functions of the number of frames used in RPCA reconstruction. Experiments using stationary and flowing microbubbles in tissue-mimicking phantoms were used to validate the method. Additionally, the method was applied to data from ten transgenic mouse models of breast cancer development, injected with B7-H3-targeted microbubbles, and two mice injected with non-targeted microbubbles. The results showed that RPCA using 20 frames achieved a Dice score of 0.95 and a computation time of 0.2 seconds, indicating that 20 frames is \textcolor{black}{potentially} suitable for real-time implementation. On in vivo data, RPCA using 20 frames achieved a Dice score of 0.82 with DTE, indicating good agreement between the two, given the limitations of each method.
\end{abstract}


\section{Introduction}
\label{sec:intro}

Ultrasound molecular imaging (UMI) is an advanced imaging modality that uses microscopic gas-filled bubbles as contrast agents to improve the detection and visualization of biological structures and processes. These microbubbles, typically 1–5 $\mu m$ in size, efficiently enhance ultrasound wave reflections, enabling high-contrast imaging. Many disease processes, including precursor lesions and invasive cancer ~\cite{klibanov2005ligand, deshpande2010molecular}, \textcolor{black}{can be} defined by the expression of specific molecules on the endothelial surface. These molecules can be targeted using special ligands attached to the microbubble shells. Through the ligand, the microbubble is bound to the diseased tissue, enabling imaging of the non-physical or molecular properties of the tissue.

A significant challenge in UMI lies in distinguishing the signal from tissue-bound targeted microbubbles from background interference, including \textcolor{black}{tissue-leakage} signals, noise, and free-floating microbubbles. Contrast enhanced ultrasound (CEUS), such as pulse inversion and amplitude modulation, ~\cite{eckersley2005optimising, phillips2001contrast, collado2024nondestructive}, \textcolor{black}{cannot distinguish tissue leakage signals from bubble signals, and it does not separate the molecular signals of bound bubbles from background noise, including free-floating bubbles.} Differential targeted enhancement (DTE) \textcolor{black}{was proposed to overcome the challenges in CEUS imaging}~\cite{lindner2001ultrasound, dayton2007molecular}. It involves using a high-intensity ultrasound pulse to burst the microbubbles, which enables a comparison between pre- and post-burst \textcolor{black}{CEUS} images~\cite{abou2015ultrasound}. However, this method has notable limitations: it destroys the contrast agents, thereby leaving ``holes'' in subsequent images; cannot be performed in real-time because it requires collecting and averaging multiple ultrasound images before and after the destruction (burst) of microbubbles; and it necessitates precise tumor localization, as it captures images from a single field of view, which restricts the broader applicability of UMI~\cite{hyun2020nondestructive}.

Previous studies employed various techniques \textcolor{black}{to overcome the limitations of DTE imaging}, including dwell-time imaging~\cite{pysz2012fast}, and minimum intensity projection~\cite{daeichin2015quantification}, to detect bound microbubbles through temporal analysis. Additionally, singular value decomposition (SVD) based methods have been utilized~\cite{mauldin2012real, herbst2019validation, collado2024nondestructive}. SVD and \textcolor{black}{principle component analysis} (PCA) algorithms are closely related to least squares procedures in their mathematical formulation, making them highly sensitive to outliers. In extreme cases, outliers can distort the computed principal components and singular vectors. The Robust Principal Component Analysis (RPCA) models data as the sum of a low-rank matrix, representing the underlying clean data, and a sparse matrix, which captures outliers and noise. This approach is more robust to outliers compared to SVD and PCA. 

RPCA has been extensively applied to various statistical tasks in computer science, aiming to identify robust principal components. Depending on the specific application, the emphasis may be placed on either the low-rank component or the sparse component. RPCA has been utilized in ultrasound localization microscopy (ULM) to enhance clutter filtering and noise suppression. The ULM approach focuses on extracting moving microbubbles and reconstructing vessels by leveraging the sparse matrix component of RPCA~\cite{xu2021robust}. We recently introduced the application of the RPCA algorithm in UMI to localize stationary targeted microbubbles bound to cancer biomarkers, while distinguishing free-floating microbubbles and provided some preliminary results~\cite{hashemi2024enhancing}. 

In this work, \textcolor{black}{we extend our work in \cite{hashemi2024enhancing} to implement a fast GPU-based version and improve our analysis by validating} RPCA filtering on simulated and experimental-phantom ultrasound data. The method is demonstrated on in vivo data from transgenic mouse models of breast cancer where B7-H3 targeted microbubbles are injected via the tail. The results of the proposed method are compared with the DTE imaging, and its potential for real-time implementation is analyzed.

\section{Methods}
\label{sec:methods}

\subsection{Overview of Proposed Robust Principle Component Analysis (RPCA) Algorithm}
For an ultrasound time-series consisting of $N_t$ image frames collected sequentially over time, a two-dimensional spatiotemporal matrix $X$ is constructed by reshaping each frame into a vector and arranging these vectors as the columns of $X$. Given that each ultrasound frame has dimensions $m \times n$, the resulting matrix has a size of $mn \times N_t$.
The echo data in $X$ is modeled as a combination of signals originating from bound microbubbles, free microbubbles, and noise. Bound microbubbles, along with possible tissue leakage, exhibits significant spatiotemporal coherence, and are modeled via a low-rank matrix $L$\textcolor{black}{, as high spatiotemporal coherence often implies redundancy and linear dependence among columns, resulting in a low-rank representation}. Conversely, the movement of free microbubbles, along with thermal noise, exhibit low \textcolor{black}{spatiotemporal} coherence and are modeled via a sparse matrix $S$\textcolor{black}{, where the columns are sparse due to the transient and unstructured nature of these components.} Consequently, the echo data can be expressed as
\begin{equation*}
X = L + S. \tag{1}
\end{equation*}
It has been shown theoretically~\cite{candes2011robust} that Principal Component Pursuit (PCP), a convex optimization technique, can effectively \textcolor{black}{decompose $X$ into its} low-rank and sparse components by solving the optimization problem
\begin{gather*}
\text{minimize} \ \Vert L\Vert_{\ast} + \lambda\Vert S\Vert_{1} \tag{2} \\
\text{subject to} \ L + S = X
\end{gather*}
where $\Vert L\Vert_{\ast} = \sum_{i} \sigma_i(L)$ is the nuclear norm of $L$, representing the sum of its singular values, and $\Vert S\Vert_{1} = \sum_{ij} \lvert S_{ij} \rvert$ is the $l_1$-norm of $S$, representing the sum of the absolute values of its elements. The parameter $\lambda$ serves as a regularization term, balancing the trade-off between the low-rank approximation and the sparse component in the matrix decomposition.

A widely used approach to \textcolor{black}{solve PCP} is the Alternating Direction Method of Multipliers (ADMM).
To optimize equation (2), ADMM forms a new objective function, known as the augmented Lagrangian function ($\mathcal{L}$), given by:
\begin{align*}
\mathcal{L}(L, S, Y) = &\ \Vert L\Vert_{\ast} + \lambda\Vert S\Vert_{1} + \langle Y, X - L - S\rangle \\
& + \frac{\mu}{2}\Vert X - L - S\Vert_{F}^{2} \tag{3}
\end{align*}
This objective function includes the original objective function from equation (2), along with \textcolor{black}{a linear term $\langle Y, X - L - S \rangle$ representing the inner product between the Lagrange multiplier $Y$ and the constraint, and a quadratic penalty term $\Vert X - L - S\Vert_{F}^{2}$, weighted by the parameter $\mu$. Both terms help enforce the constraint as the iterations progress~\cite{candes2011robust}.} ADMM solves equation (3) iteratively through \textcolor{black}{repeating three} steps:
\begin{enumerate}
    \item Minimize $\mathcal{L}$ with respect to $L$ (fixing $S$ and $Y$):
    It has been shown~\cite{candes2011robust} that minimizing the augmented Lagrangian in Equation (3) with respect to $L$ is equivalent to performing \textcolor{black}{thresholded singular value decomposition (SVD) with a threshold of $\tau = \frac{1}{\mu}$, where the largest singular values are retained in $L$ while the smaller ones are set to zero, allowing $L$ to capture the dominant structures in the data.} \textcolor{black}{This can be defined as:
    \begin{equation*}\underset{L}{\min}\;\mathcal{L}(L, S, Y)=D_{\frac{1}{\mu}}(X-S+\mu^{-1}Y) \tag{4}\end{equation*}
    where the singular value thresholding operator for any matrix $M$ is given by $D_{\tau}(M) = U\Lambda_{\tau} V^{*}$, with  $\Lambda_{\tau}=$~max$(\Lambda-\tau ,0)$ applying the threshold $\tau = \frac{1}{\mu}$ to the singular values of $M$. Here, $M = U\Lambda V^{*}$ represents the SVD of $M$, where $\Lambda$ is the diagonal matrix of singular values.}

    \item Minimize $\mathcal{L}$ with respect to $S$ (fixing $L$ and $Y$): It has been shown~\cite{candes2011robust} that this minimization problem is \textcolor{black}{equivalent to element-wise thresholding of the residual, $X - L$, meaning that small-magnitude elements are set to zero while large-magnitude elements remain. In this way, $S$ can capture the outliers.}
    \textcolor{black}{This has been defined by the shrinkage operator:
    \begin{equation*}
    S=\mathrm{shrink}_{\frac{\lambda}{\mu}}(X-L+\mu^{-1}Y) \tag{5}
    \end{equation*}
    where the shrinkage operator is defined as $\mathrm{shrink}_{\gamma}[x] = \mathrm{sgn}(x)\max(\lvert x \rvert - \gamma, 0)$ where $x$ represents an individual element of the matrix $X-L+\mu^{-1}Y$. 
    This means that for each entry $x_{ij}$ in $X-L+\mu^{-1}Y$, it is thresholded according to $\gamma=\frac{\lambda}{\mu}$: if $x_{ij}$ is greater than $\gamma$, it is reduced by $\gamma$; otherwise, it is set to zero. The output will be the updated sparse matrix $S$, which contains the thresholded values. }
    
    \item Update Lagrangian multiplier $Y$: In practice, \textcolor{black}{since the updates for $L$ and $S$ in the first two steps are performed separately, inconsistencies may arise such that $L+S$ does not exactly equal $X$. The Lagrange multiplier Y helps gradually correct these discrepancies.} The update for Y is given by:
    \begin{equation*} Y_{k+1}=Y_{k}+\mu(X-L_{k}-S_{k}) \tag{6}\end{equation*}
    In this work, we choose $\mu = \frac{n_1 n_2}{4\Vert X \Vert_{1}}$ where $n_1$ and $n_2$ are the dimensions of the data matrix $X$, as suggested in \cite{yuan2009sparse}. We terminate the algorithm when $\Vert X - L - S\Vert_{F}<\delta \Vert X\Vert_{F}$, with $\delta = 10^{-9}$. The subscript \( F \) in \( \Vert \cdot \Vert_F \) refers to the Frobenius norm of a matrix where, for matrix \( X \), is defined as:
$\Vert X \Vert_F = \sqrt{\sum_{i,j} |X_{i,j}|^2}$.

\end{enumerate}

\subsection{GPU Implementation}

We have \textcolor{black}{adapted} a GPU-based implementation of the RPCA method using PyTorch to efficiently decompose ultrasound molecular imaging data into low-rank and sparse components\textcolor{black}{~\cite{Reinhold2019cuda}}. This implementation leverages CUDA acceleration to handle large datasets while maintaining computational efficiency. 
The input RF data is normalized and reshaped \textcolor{black}{to form the data matrix $X$ before decomposition}. 
The regularization parameter $\lambda$ is set \textcolor{black}{as explained earlier} and $\mu$ is initialized as $\mu = \frac{n_1 n_2}{4\Vert X \Vert_{1}}$, where $n_1$ and $n_2$ are the dimensions of each RF frame. We initialize and update $L$, $S$, and the dual variable $Y$ iteratively until convergence, using the Frobenius norm as the stopping criterion.

\subsection{Simulation}
\subsubsection{Simulations for Microbubble Imaging}
Ten simulations were performed using a random number and location of stationary and moving microbubbles in the image plane. \textcolor{black}{For the total bound and free microbubbles}, a random number between 800 and 1600 \textcolor{black}{microbubbles} were selected and randomly spread over a $6\times5~{mm}^2$, resulting in an average density of 27 to 54 $microbubbles/{mm}^2$.
The simulations were performed in \textcolor{black}{Field II~\cite{jensen1997field,
jensen1992calculation}}, using the Marmottant model\textcolor{black}{~\cite{marmottant2005model}} to represent the nonlinear frequency response of microbubbles. Realistic microbubble responses to incident ultrasonic pulses were simulated using BubbleSim~\cite{hoff2001acoustic}, and the resulting acoustic backscatter was \textcolor{black}{modeled using} a simulated L12-3v ultrasound transducer \textcolor{black}{in} Field II. 
The tissue simulations are performed separately from the microbubble simulations, and their RF signals are then combined using superposition. Thermal noise is subsequently added at -12 dB relative to the \textcolor{black}{root-mean-square} (RMS) of the B-mode signal. In the case of contrast mode, only the RF signals of stationary and moving microbubbles are combined with the thermal noise. 
For each simulation, a time series of harmonic images consisting of 60 frames was generated and served as the input for the RPCA algorithm. Angled plane wave transmissions from 25 different angles were simulated modeling a Verasonics L12-3v transducer (Kirkland, WA, USA) and coherently compounded via delay-and-sum beamforming to produce synthetically focused radiofrequency (RF) data. \textcolor{black}{Table~\ref{tab:simParam} presents the parameters used in the Field II simulation and BubbleSim software}.

\begin{table}[h]
    \centering
    \begin{tabular}{|c|c|}
        \hline
        Parameter & Value \\ 
        \hline
        Transducer   & Linear (L12-3v)   \\
        Number of elements   & $128$   \\
        Pitch [mm]  & $0.2$   \\
        Center frequency [MHz]  & $8$   \\ 
        Bandwidth [MHz]  & $[3,12]$   \\ 
        TX angular range   & $[-5^{\circ}, 5^{\circ}]$   \\ 
        TX angle spacing   & $0.4^{\circ}$   \\
        \hline
        Bubble radius [$\mu m$]   & $1$   \\
        Internal equilibrium pressure [$Pa$]   & $1.013\times10^5$   \\
        Density of surrounding liquid [$kg/m^3$]   & $1000$   \\
        Viscosity of the liquid [$Pas$]   & $10^{-3}$   \\
        Speed of sound in the liquid [$m/s$]   & $1500$   \\
        Thermal conductivity of the gas [$W/mK$]   & $26.2\times10^{-3}$   \\
        Density of gas [$kg/m^3$]   & $1.161$   \\
        Heat capacity of the gas [$J/kgK$]   & $1.007\times10^3$   \\
        Adiabatic constant of the gas   & $1.4$   \\
        \hline
    \end{tabular}
    \caption{Simulation parameters.}
    \label{tab:simParam}
\end{table}

\subsubsection{Modeling of microbubble motion}
The motion of microbubbles is governed by Newton’s second law, with forces such as fluid drag, acoustic radiation force, and buoyancy potentially influencing their trajectories. However, in our model, bubble motion is simplified by considering only random accelerations and velocities. 
Each bubble is assigned an initial velocity amplitude, $v_0$ which is randomly selected from a uniform distribution in the range of $[0.01, 2]~cm/s$, ensuring realistic microbubble speeds. The velocity components in the axial and lateral directions are determined by multiplying $v_0$ with the cosine and sine of a randomly chosen angle within $[-\pi,\pi]$, allowing bubbles to start in random directions. To introduce variability in bubble trajectories, a random acceleration within the range of $[1,10]~cm/s^2$ is applied at each time step. The acceleration components are obtained by multiplying this value by the cosine and sine of a random angle within $[-\pi,\pi]$, ensuring a reasonable spread in different directions.

At each time step, velocities and new positions are updated using the following equations in the axial and lateral directions separately: $v(t) = v(t-1)+a\cdot dt$, and $x(t) = x(t-1)+v(t)\cdot dt$. Based on this model, the moving bubbles start at random positions and directions and follow curved paths due to random accelerations applied over time. Some bubbles exhibit more deviation, while others have relatively straight trajectories depending on their initial conditions and variations in acceleration magnitude and direction. The presence of a time-dependent velocity update suggests a gradual shift in trajectory rather than sudden changes.

\begin{figure*}
    \centering
    \begin{subfigure}{0.205\textwidth}
        \centering
        \includegraphics[width=\textwidth]{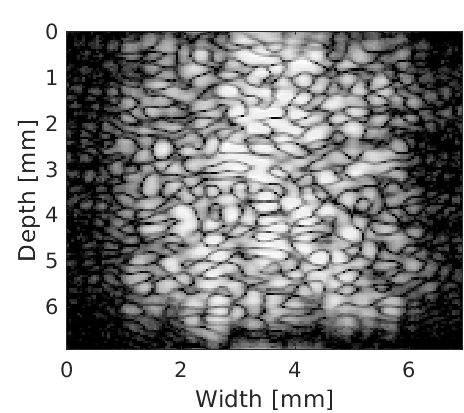}
        \caption{Contrast image}
        \label{fig:a}
    \end{subfigure}
    \begin{subfigure}{0.18\textwidth}
        \centering
        \includegraphics[width=\textwidth]{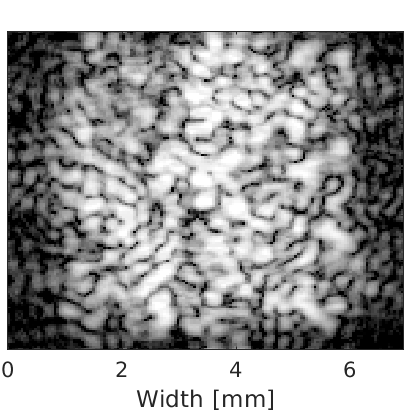}
        \caption{L (RPCA)}
        \label{fig:b}
    \end{subfigure}
        \begin{subfigure}{0.18\textwidth}
        \centering
        \includegraphics[width=\textwidth]{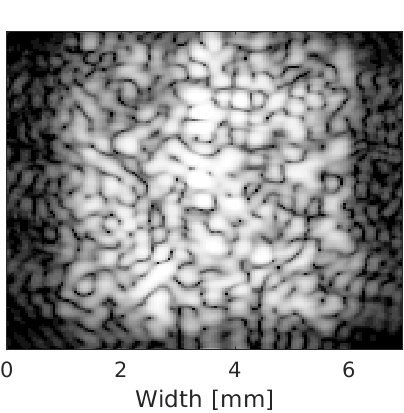}
        \caption{L (ground-truth)}
        \label{fig:b}
    \end{subfigure}
    \begin{subfigure}{0.18\textwidth}
        \centering
        \includegraphics[width=\textwidth]{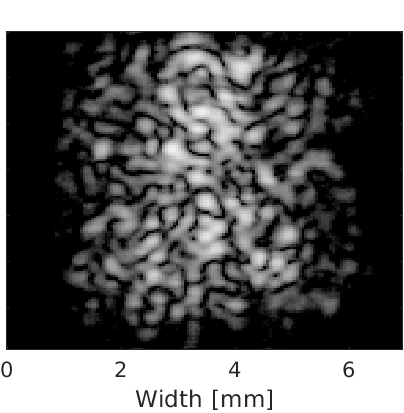}
        \caption{S (RPCA)}
        \label{fig:b}
    \end{subfigure}
    \begin{subfigure}{0.214\textwidth}
        \centering
        \includegraphics[width=\textwidth]{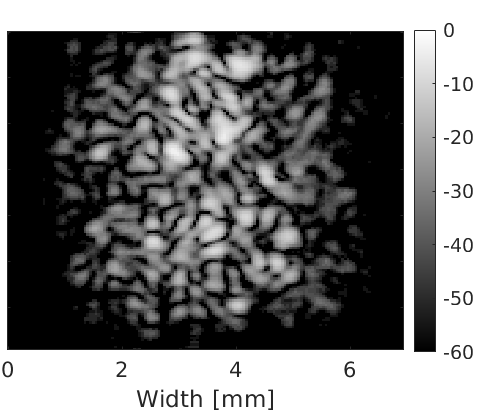}
        \caption{S (ground-truth)}
        \label{fig:b}
    \end{subfigure}
    \caption{Representative image from a time-series simulation of stationary and moving microbubbles. The original contrast image in (a) is decomposed into low-rank and sparse matrices, shown in (b) and (d), which represent the stationary and moving microbubbles, respectively. Their corresponding ground-truth images are shown in (c) and (e).}
    \label{fig:sim_images}
\end{figure*}

\begin{figure}
    \centering
    \begin{subfigure}{0.4\textwidth}
        \centering
        \includegraphics[width=\textwidth]{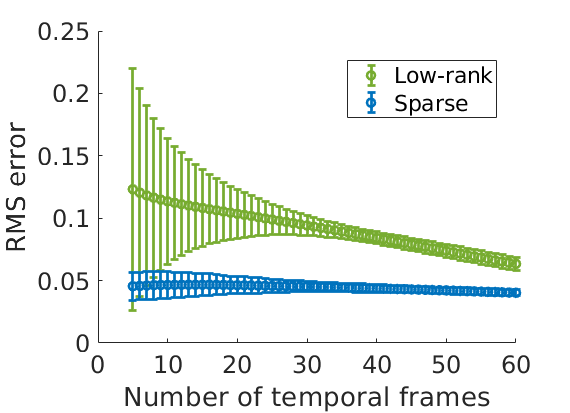}
        \caption{RMSE}
        \label{fig:a}
    \end{subfigure}
    \begin{subfigure}{0.4\textwidth}
        \centering
        \includegraphics[width=\textwidth]{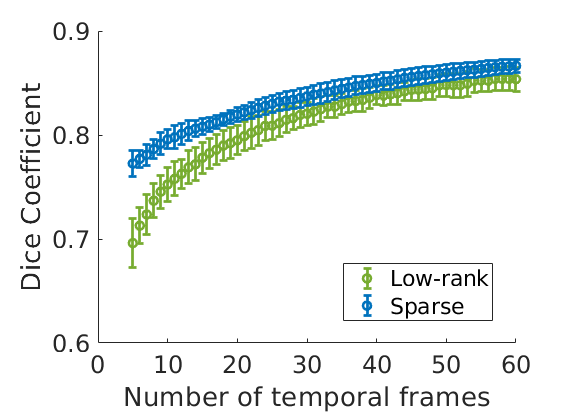}
        \caption{Dice coefficient}
        \label{fig:b}
    \end{subfigure}
    \caption{The RMS error (a) and Dice coefficient (b) as a function of the number of frames over time were calculated between the low-rank and sparse matrices and their corresponding ground-truth images.}
    \label{fig:sim_measurements}
\end{figure}

\subsection{Phantom Experiment}

An agar phantom with a 2 mm flow channel was created to mimic a vessel within tissue. The phantom consisted of 2\% agarose and 2\% graphite \textcolor{black}{with one end of the phantom closed}. To simulate stationary bubbles, we created custom non-targeted microbubbles according to the protocol described in~\cite{abou2021horizon}, \textcolor{black}{diluted the microbubbles with saline} and mixed them with ultrasound gel to enhance their stability, and then injected them into the \textcolor{black}{open end of the channel}. After injection, the \textcolor{black}{open} side of the channel was also \textcolor{black}{closed}, and the microbubbles were allowed to settle and become stationary. Time-series beamformed RF data consisting of 20 frames were captured using an ACUSON S3000 ultrasound scanner (Siemens Medical Solutions, Mountain View, CA, USA) in the Contrast Mode at a frame rate of 20 frames/s with a MI of 0.07. The conventional focused transmit data were acquired using a 9L4 linear transducer at 5~MHz center frequency and in the Research Mode\cite{brunke2007ultrasound}.
\textcolor{black}{To simulate moving microbubbles, they were first diluted in saline. A peristaltic pump (ISMATEC, IDEX Corporation, IL, USA) and its tubing were connected to both ends of the phantom channel to create a closed-loop system for continuous microbubble circulation. The maximum system flow rate of $0.148~mL/min$ was used, corresponding to a velocity of approximately $0.79~mm/s$ in tubing with a $2~mm$ diameter. This value closely approximates the blood flow velocity in human and mouse capillaries.} 


\subsection{In Vivo Data Collection}
All animal experiments were approved by the Institutional Administrative Panel on Laboratory Animal Care at Stanford University under protocol 14686 \textcolor{black}{and adhered to the ARRIVAL 2.0 guidelines}. 
Ten transgenic mouse models of breast cancer development~\cite{lin2003progression} were used to collect in vivo UMI data.
Imaging was performed using B7-H3 targeted microbubbles. B7-H3 is a glyco-protein belonging to the B7 family of immune checkpoint molecules, which plays a role in regulating immune responses. Its expression is elevated in several types of cancer, including breast cancer, where it has proven to be an effective biomarker for differentiating between normal, benign, precursor, and malignant breast pathologies using UMI~\cite{bachawal2015breast}.

\subsubsection{Preparation of B7-H3 Targeted Microbubbles}
B7-H3 targeted microbubbles were prepared using \textcolor{black}{functionalized} microbubble formulations purchased in liquid form \textcolor{black}{(USphere Labeler LA, Trust Biosonics, Tapei)} and stored at $4^\circ \mathrm{C}$. The biotinylated antibody targeting B7-H3 was prepared by conjugating monoclonal antibodies (mAbs) with NHS-PEG4-Biotin (ThermoFisher Scientific) at a 1:2 molar ratio. NHS-PEG4-Biotin was dissolved in phosphate-buffered saline \textcolor{black}(PBS) and incubated with mAbs (1 mg/mL in UltraPure Water) at room temperature for 30 minutes. Unconjugated biotin was removed using a 7 kDa molecular weight cut-off Zeba spin column. For \textcolor{black}{targeted microbubble} preparation, $1 \times 10^8$ microbubbles/mL/vial were mixed with $10~\mu g$ of the biotinylated B7-H3 mAb, incubated for 1 hour at room temperature, and washed three times with PBS using a microcentrifuge at $300~g$ for 3 minutes. The topmost floating \textcolor{black}{microbubbles} were carefully separated and resuspended in fresh PBS. Nontargeted microbubbles made from the same source as above but without the B7-H3 antibody were used in two mice as a negative control.

\subsubsection{Imaging Setup}
\textcolor{black}{B7-H3 targeted microbubbles were injected in ten mice through the mouse tail vein and allowed to circulate for 3 minutes to ensure adequate binding to the B7-H3 target and to clear some of the free microbubbles.} A Verasonics Vantage 256 research scanner with an L12-3v transducer was used to collect in vivo RF channel data. \textcolor{black}{We acquired 132 image frames, then burst the bubbles by sending a strong acoustic pulse, and acquired the same number of frames afterward.}
Each acquisition included 25 plane wave steering angles uniformly distributed over $[-5^\circ,5^\circ]$ angular range.
For each steering angle, a 10~MHz pulse was transmitted to reconstruct the B-mode image, while two pulses at a fundamental frequency of 5~MHz with opposite polarities were transmitted \textcolor{black}{for CEUS imaging}. The received signals from the 5~MHz pulses were summed to perform pulse inversion \textcolor{black}{CEUS} imaging, targeting the second harmonic at 10~MHz. 
To generate DTE images, the average of the \textcolor{black}{CEUS} frames after the burst is subtracted from a moving average of every 20 frames before the burst, creating corresponding DTE time-series.


\begin{figure*}
    \centering
    \includegraphics[width=\textwidth]{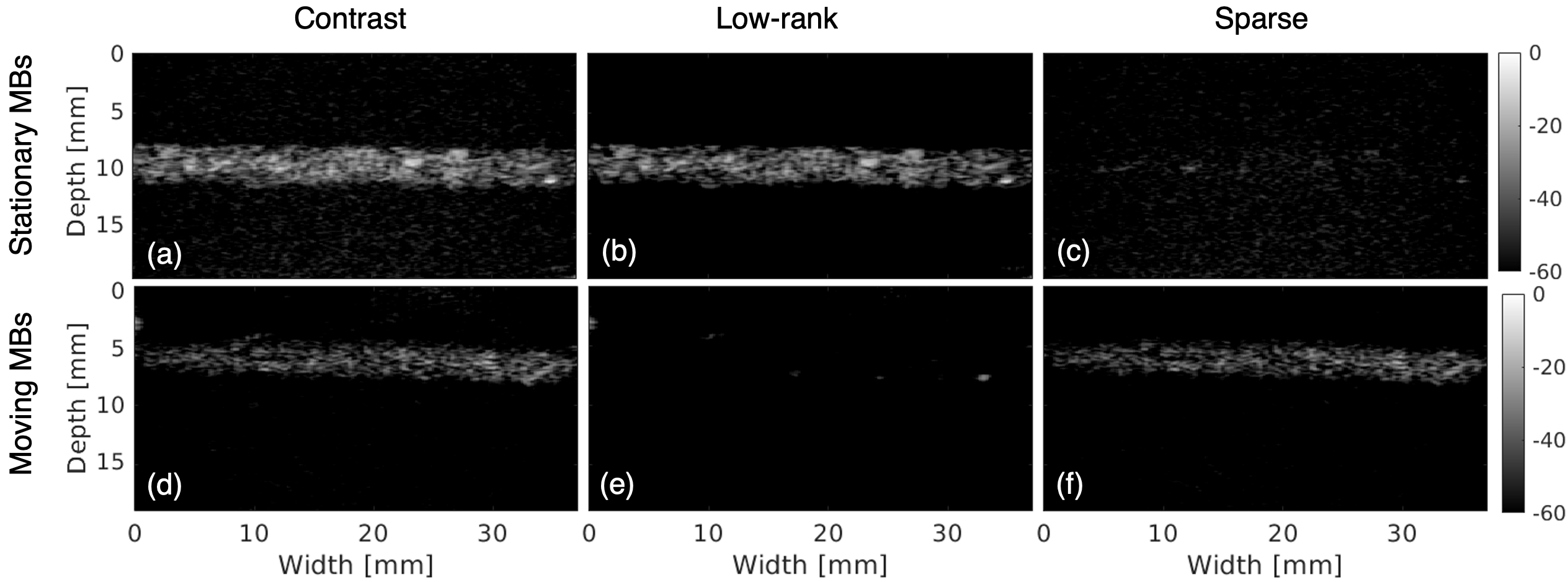}
    \caption{Phantom experiments showing contrast images of (a-c) stationary and (d-f) moving microbubbles. Top row: (a) Contrast image of nontargeted microubbles in gelatin injected into a channel. The contrast image was decomposed into a (b) low rank and (c) sparse image. The stationary microbubbles appear in the low rank image with little or no bubble signal in the sparse image, as expected. (d) Contrast image of moving microbubbles. The contrast image was decomposed into a (e) low rank and (f) sparse image. No signal appears in the low rank image and the moving microbubbles only appear in the sparse image, as expected.}
    \label{fig:phantomExp}
\end{figure*}

\subsubsection{Applying the RPCA method}
For the in vivo data, 20, 60, and 132 frames were used to construct the data matrix $X$ \textcolor{black}{and then solve for the low-rank and sparse matrices $L$ and $S$}. The RPCA filter applied to 132 frames \textcolor{black}{was used as} a reference, as using more frames improves accuracy \textcolor{black}{of the low rank image} at the cost of longer running times. \textcolor{black}{For this reference, to further filter out respiratory motion, we excluded frames with a correlation value below 0.8 relative to their neighboring frames, and a temporal moving average filter with a window of 20 frames was applied to the data.}

\subsection{Evaluation Metrics}
For the simulation data, the root-mean-square error (RMSE) \textcolor{black}{of the RF data} and Dice coefficient of the \textcolor{black}{CEUS} images are used to compare the estimated low-rank and sparse matrices with the ground truth images. The \textcolor{black}{CEUS} images were converted to binarized versions after histogram matching, and using a threshold based on the average intensity of the pixels at the bubble locations in the
ground-truth image.
For the in vivo data, the dice coefficient is used to compare the results with binary versions of the reference and the DTE images. To convert the images to binary, the average intensity of the samples \textcolor{black}{within a $3\times3~mm$ window} in the microbubble region was used as the threshold value. For images of size $m \times n$ and $t$ frames over time, these metrics are defined as follows:
\begin{itemize}
    \item Root-mean-square error:
    \begin{equation*} \textrm {RMSE} = \sqrt{\frac{1}{mnk} \sum_{t=1}^{k} \sum_{j=1}^{n} \sum_{i=1}^{m} \left( \hat{c}_{ijt} - c_{ijt} \right)^2},\tag{7}
    \end{equation*} 
where $c$ and $\hat{c}$ represent the ground truth and estimated values at image sample $(i,j)$ and frame $k$ over time, calculated separately for the low-rank and sparse matrices.
    \item Dice Similarity Coefficient:
    \begin{equation*}
    \text{Dice}(C, \hat{C}) = \frac{2|C \cap \hat{C}|}{|C| + |\hat{C}|},\tag{8}
    \end{equation*} 
    where $C$ and $\hat{C}$ are the estimated and ground-truth (or reference) binary images.
\end{itemize} 

The Dice coefficient is a commonly used metric for evaluating the overlap between two segmented regions, but it is sensitive to class imbalance. In the case of non-targeted microbubbles, the number of microbubbles is significantly lower than the background samples, creating a highly imbalanced dataset where the background dominates. When one class (background) is much larger than the other (microbubbles), the Dice coefficient may not accurately compare DTE and low-rank images. Therefore, the Dice coefficient is not measured for the in vivo cases with non-targeted microbubbles.

\section{Results}
\label{sec:results} 

\subsection{Simulation}
An example frame from the time series of one of the simulated samples is shown in Figure~\ref{fig:sim_images}, where 20 frames were used to construct the data matrix. The original contrast image in (a) is decomposed into low-rank and sparse matrices, as shown in (b) and (d). The low-rank matrix represents the stationary microbubbles, while the sparse matrix captures the moving microbubbles. Their corresponding ground-truth images are shown in (c) and (e). The Dice coefficients are 0.82 and 0.81 for the low-rank and sparse matrices, respectively, and the corresponding RMS errors are 0.14 and 0.05.

The RMS error and Dice coefficient across ten simulations are calculated and presented in Figure~\ref{fig:sim_measurements} as a function of the number of temporal frames used in the RPCA filter. For the low-rank matrix, the RMS error decreases as more frames are used, whereas for the sparse matrix, it remains relatively constant regardless of the number of frames. This is due to thermal noise, which causes the RMSE to reach a lower bound, as low-amplitude thermal noise is never fully reconstructed. The Dice coefficients were calculated between the estimated and ground-truth images, as shown in Figure~\ref{fig:sim_measurements} (c) and (d). 
Based on the RMS error and Dice coefficient plots for the low-rank matrix, using 20 frames to reconstruct the data matrix $X$ yields an RMSE of 0.1 and a Dice coefficient of 0.75.

\subsection{Phantom}
In Figure~\ref{fig:phantomExp}, the top row shows the application of RPCA filtering to stationary microbubbles in a phantom. Panel (a) shows the contrast mode image, where the gel-mixed microbubbles are visible inside the channel. A time series of 20 contrast-mode images is used in the RPCA algorithm to generate time series of low-rank and sparse images, which are shown in panels (b) and (c), respectively. \textcolor{black}{Because} most of the bubbles are stationary, they appear in the low-rank image as expected, while the sparse image mostly captures ultrasound noise. 

The second row shows the application of RPCA filtering to moving  microbubbles in a phantom. Panels (d)–(f) present the original contrast mode image, low-rank image, and sparse image, respectively. As expected, the moving microbubbles are present only in the sparse image. Movies showing the time series contrast mode images and corresponding low-rank and sparse images for these two cases are \textcolor{black}{included} in the \textcolor{black}{supplementary material}.

\begin{figure*}
    \centering
    \begin{subfigure}{0.27\textwidth}\includegraphics[height=3.7cm]{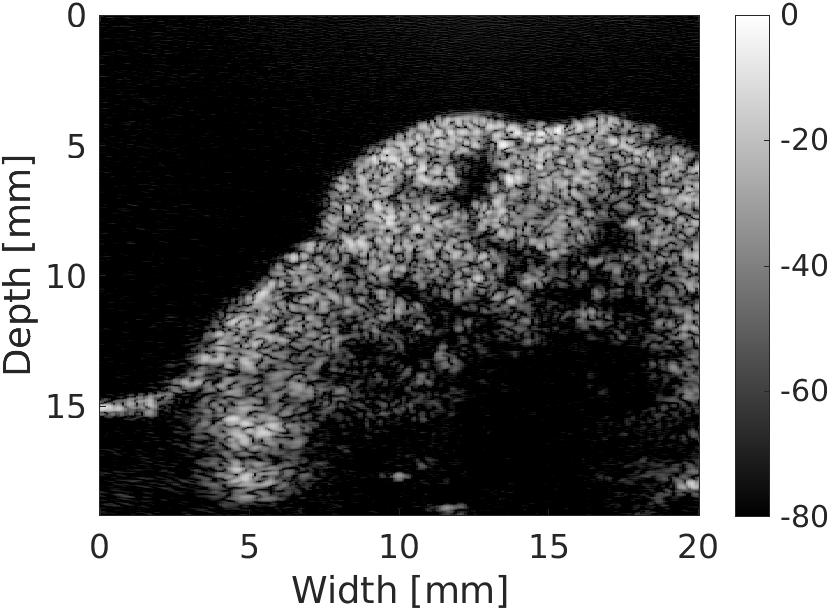}
    \caption{Contrast image}
    \end{subfigure}
    \begin{subfigure}{0.21\textwidth}\includegraphics[height=3.7cm]{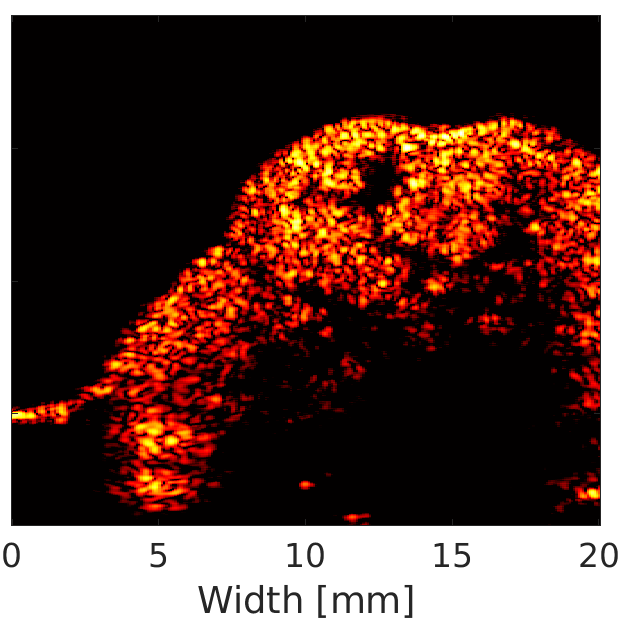}
    \caption{L (20 frames)}
    \end{subfigure}
    \begin{subfigure}{0.21\textwidth}\includegraphics[height=3.7cm]{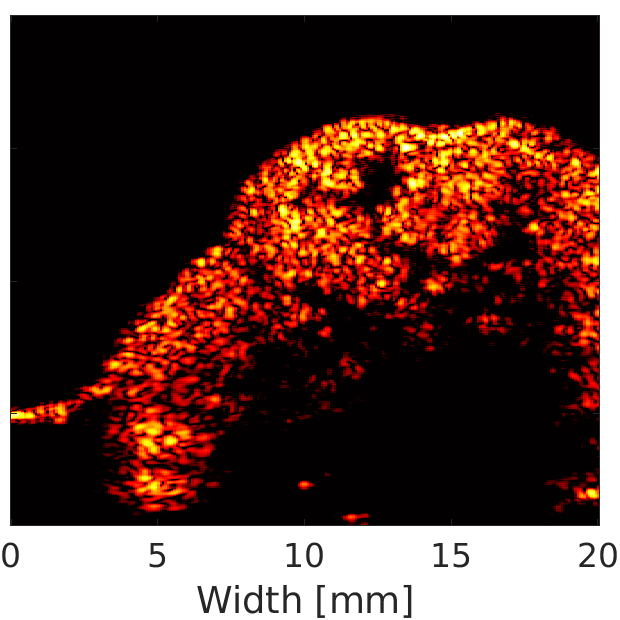}
    \caption{L (60 frames)}
    \end{subfigure}
    \begin{subfigure}{0.21\textwidth}\includegraphics[height=3.7cm]{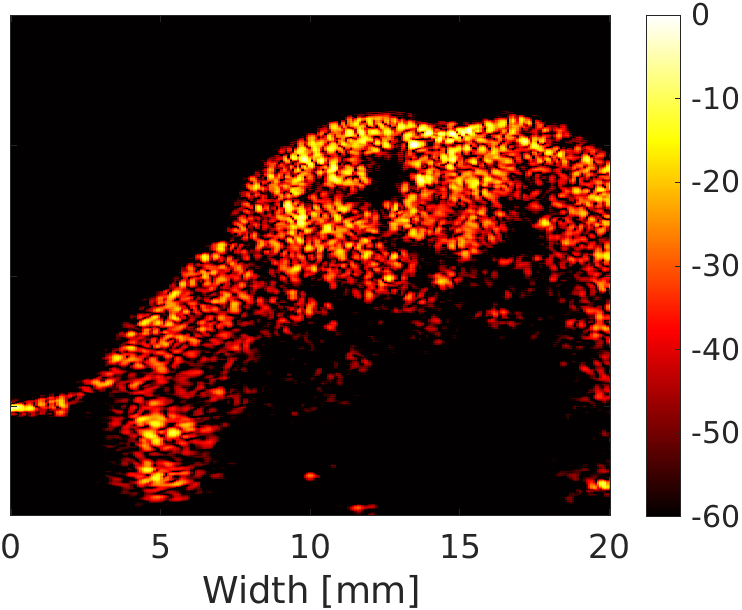}
    \caption{L (132 frames)}
    \end{subfigure}\\
    
    \vspace{0.2cm}
    
    \begin{subfigure}{0.24\textwidth}\includegraphics[height=3.7cm]{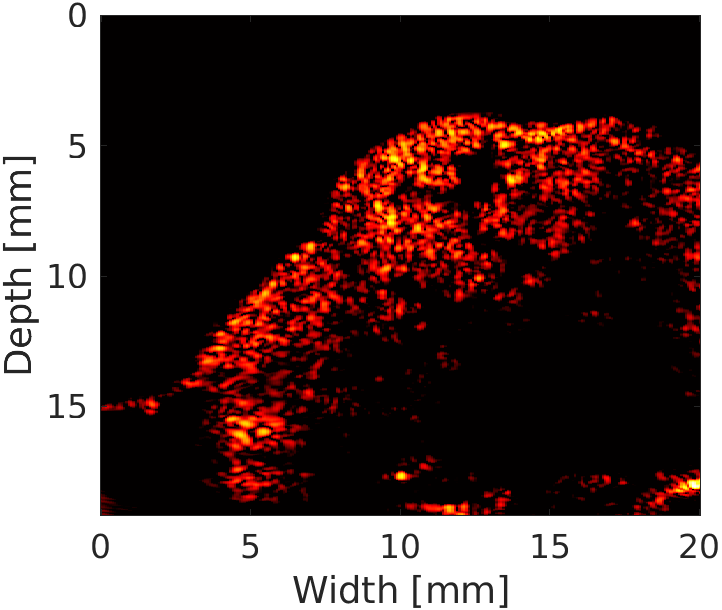}
    \caption{DTE}
    \end{subfigure}
    \begin{subfigure}{0.22\textwidth}\includegraphics[height=3.7cm]{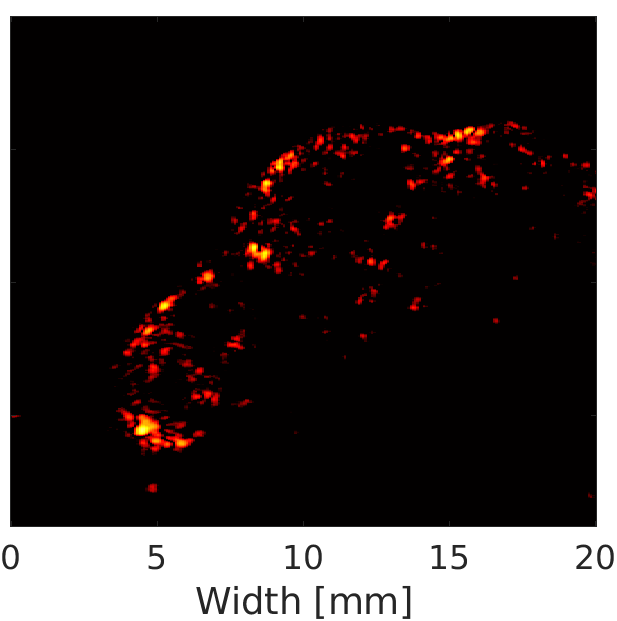}
    \caption{S (20 frames)}
    \end{subfigure}
    \begin{subfigure}{0.22\textwidth}\includegraphics[height=3.7cm]{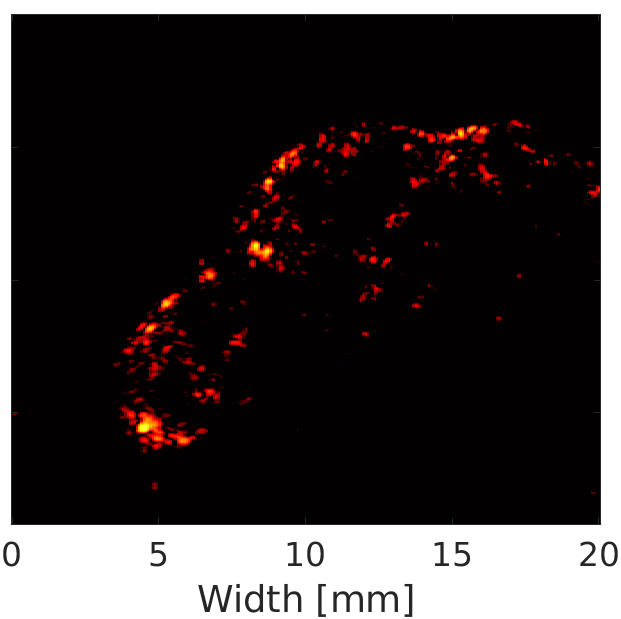}
    \caption{S (60 frames)}
    \end{subfigure}
    \begin{subfigure}{0.22\textwidth}\includegraphics[height=3.7cm]{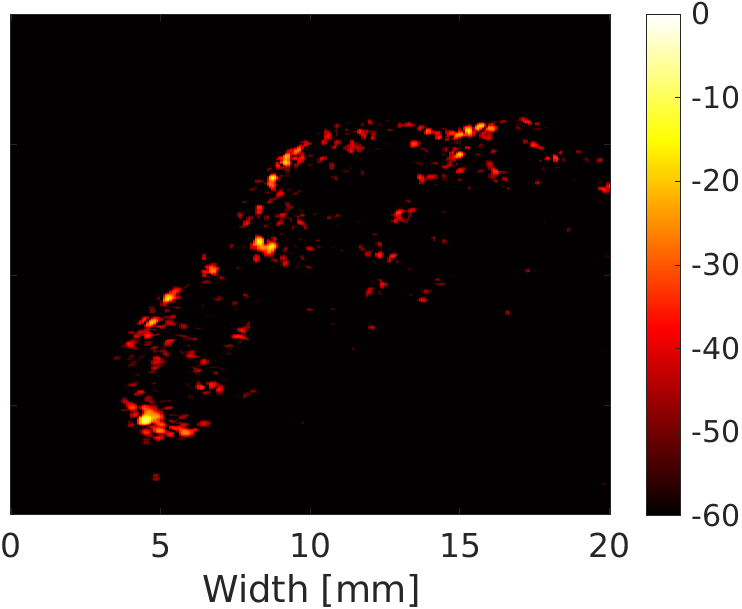}
    \caption{S (132 frames)}
    \end{subfigure}
    \caption{In vivo experiment results. (a)–(b) show the contrast image and the DTE result for bound bubbles. The low-rank images obtained from RPCA using 20, 60, and 132 frames are shown in (c)–(e). The corresponding sparse images from RPCA are shown in (f)–(h).}
    \label{fig:invivo_images}
\end{figure*}

\begin{figure}
    \centering    
    \begin{subfigure}{0.49\textwidth}
        \centering
        \includegraphics[width=\textwidth]{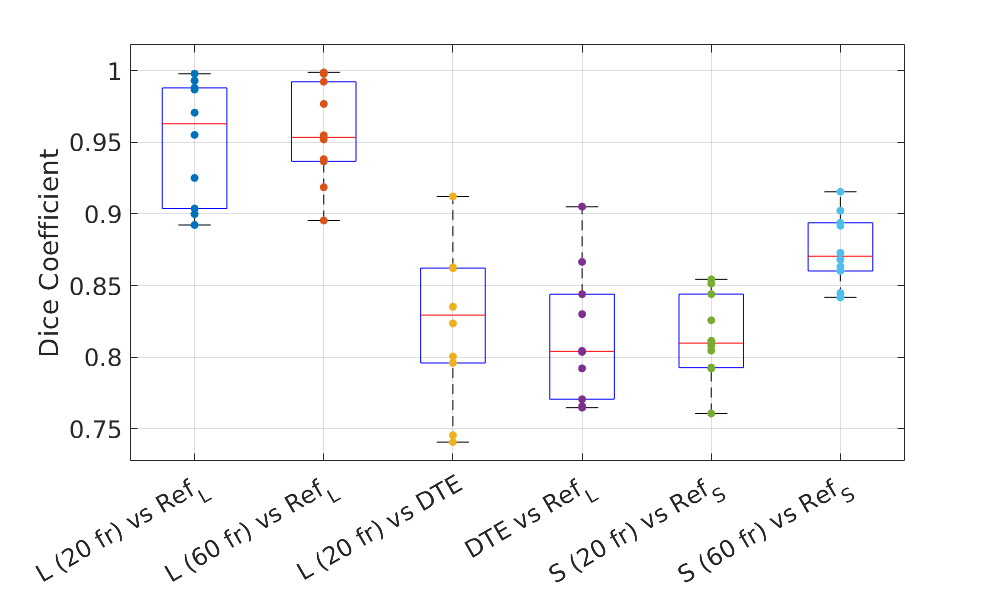}
    \end{subfigure}
    \caption{Box plot of the Dice values for the in vivo experimental results across 10 different mice for the low-rank and sparse matrices. Each data point represents the Dice coefficient calculated for an individual mouse.}
    \label{fig:invivo_dice_plots}
\end{figure}

\subsection{In Vivo}
Figure~\ref{fig:invivo_images} shows a representative example of the in vivo molecular images targeting B7-H3 in breast cancer. Panels (a)–(b) display the original contrast image and the DTE image. The low-rank images obtained from the proposed RPCA method using different numbers of frames are shown in (c)–(e). The corresponding sparse images from RPCA are shown in (f)–(h).

Using the RPCA filter on 132 frames as the reference, the average Dice coefficients across 10 mice for the RPCA results using 20 frames are $0.95\pm0.04$ for the low-rank image and $0.81\pm0.02$ for the sparse image. The average Dice coefficients for the RPCA filter using 60 frames are $0.96\pm0.03$ for the low-rank matrix and $0.87\pm0.02$ for the sparse matrix.
The average Dice coefficient between the low-rank image using 20 frames and the DTE image is $0.82 \pm 0.05$. The corresponding Dice value \textcolor{black}{between the reference} low-rank image using 132 frames and the DTE image is $0.81 \pm 0.04$. This indicates good agreement between the two methods.
\textcolor{black}{Boxplots of} results are shown in Figure~\ref{fig:invivo_dice_plots}, where each data point representing the Dice coefficient calculated for an individual mouse. \textcolor{black}{For} each box, the central mark indicates the median, and the bottom and top edges of the box indicate the 25th and 75th percentiles, respectively.

\begin{figure*}
    \centering
    \begin{subfigure}{0.27\textwidth}\includegraphics[height=3.7cm]{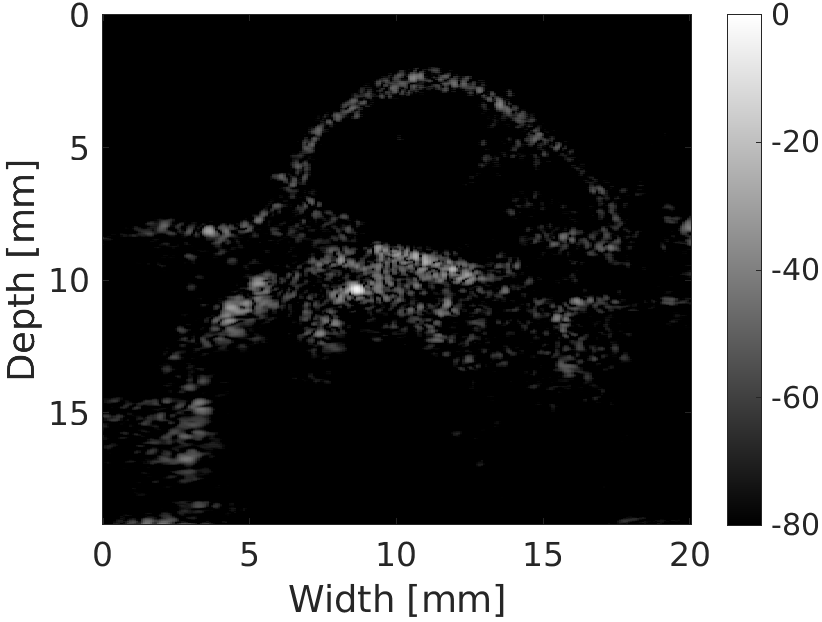}
    \caption{Contrast image}
    \end{subfigure}
    \begin{subfigure}{0.21\textwidth}\includegraphics[height=3.7cm]{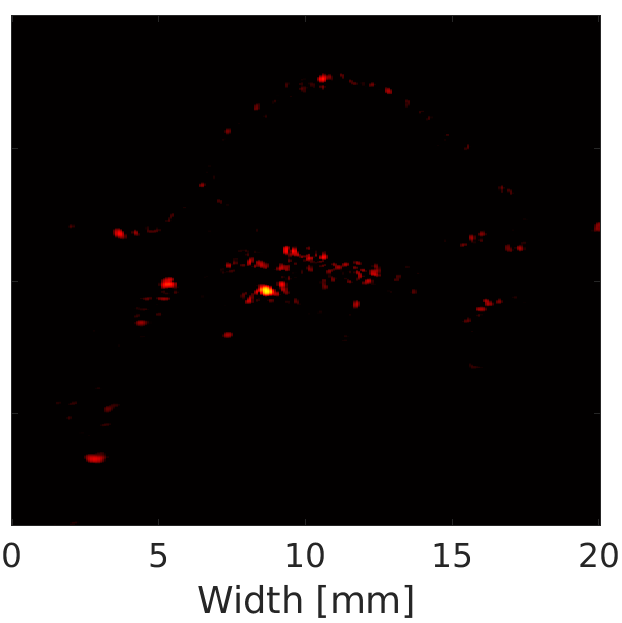}
    \caption{DTE}
    \end{subfigure}
    \begin{subfigure}{0.21\textwidth}\includegraphics[height=3.7cm]{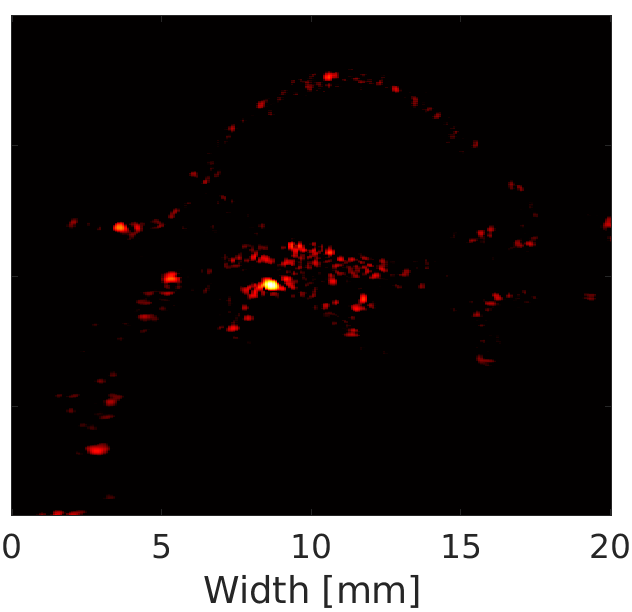}
    \caption{L}
    \end{subfigure}
    \begin{subfigure}{0.21\textwidth}\includegraphics[height=3.7cm]{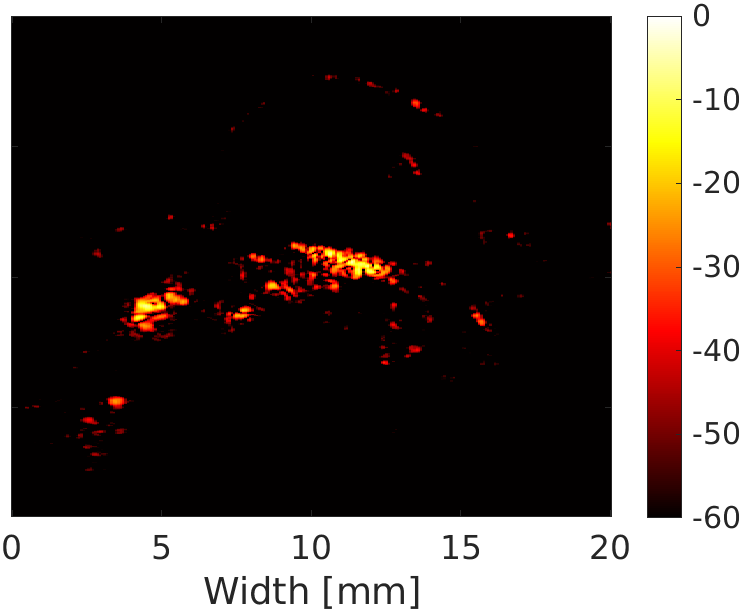}
    \caption{S}
    \end{subfigure}\\
    
    \vspace{0.2cm}
    
    \begin{subfigure}{0.27\textwidth}\includegraphics[height=3.7cm]{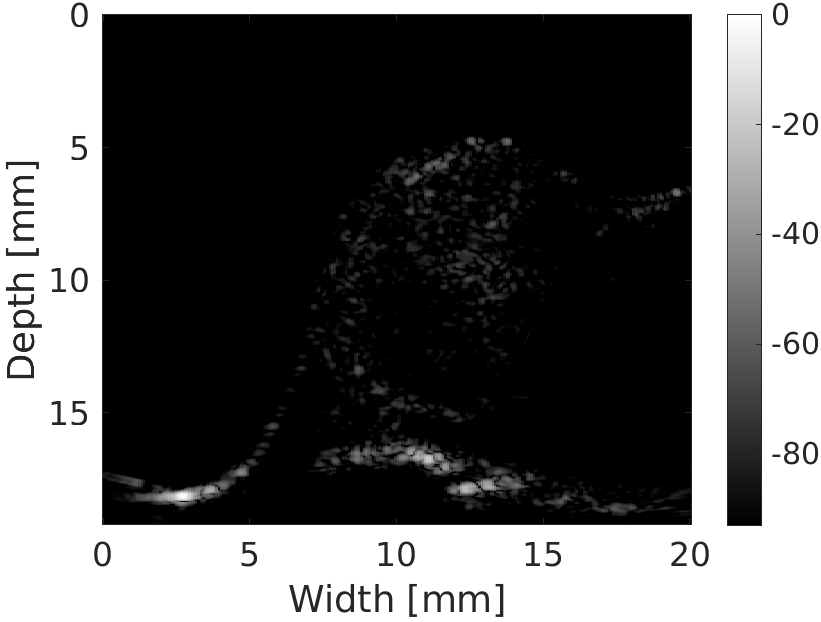}
    \caption{Contrast image}
    \end{subfigure}
    \begin{subfigure}{0.21\textwidth}\includegraphics[height=3.7cm]{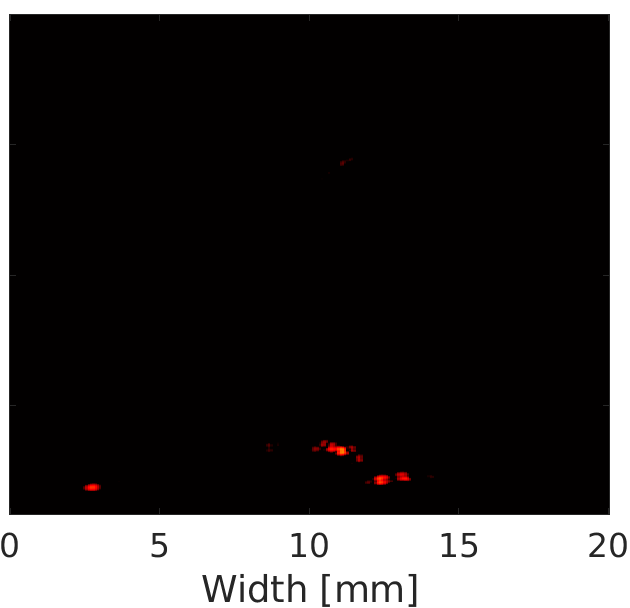}
    \caption{DTE}
    \end{subfigure}
    \begin{subfigure}{0.21\textwidth}\includegraphics[height=3.7cm]{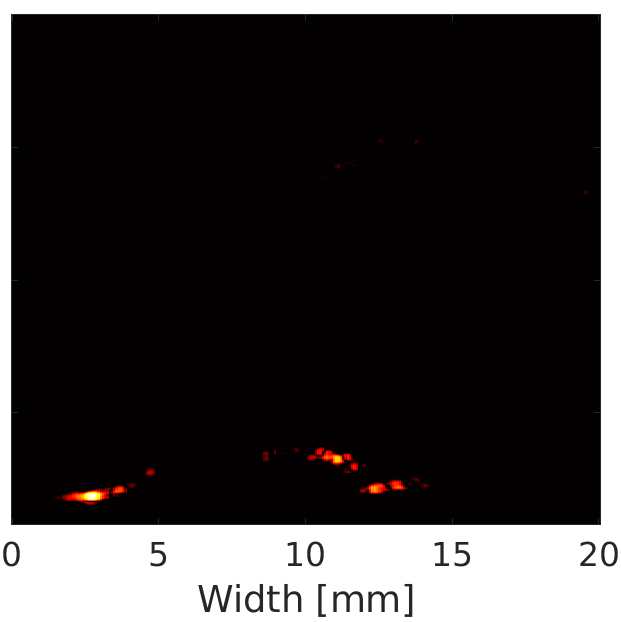}
    \caption{L}
    \end{subfigure}
    \begin{subfigure}{0.21\textwidth}\includegraphics[height=3.7cm]{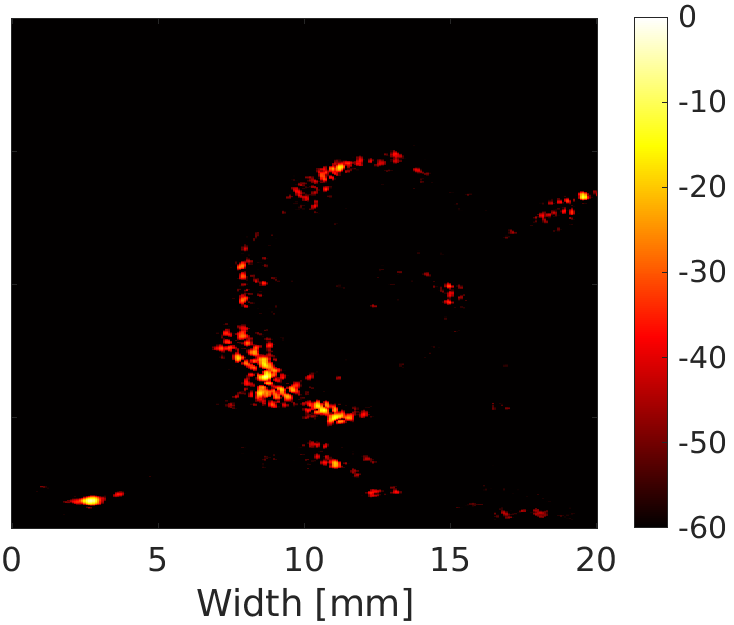}
    \caption{S}
    \end{subfigure}
    \caption{In vivo experiment results with nontargeted bubbles. The first and second rows correspond to different mice. From left to right for each mouse: the original harmonic image, the DTE image, the estimated low-rank image using RPCA, and the sparse image obtained from RPCA.}
    \label{fig:invivo_images_nonTargeted}
\end{figure*}

\begin{figure}
    \centering    
    \begin{subfigure}{0.49\textwidth}
        \centering
        \includegraphics[width=\textwidth]
        {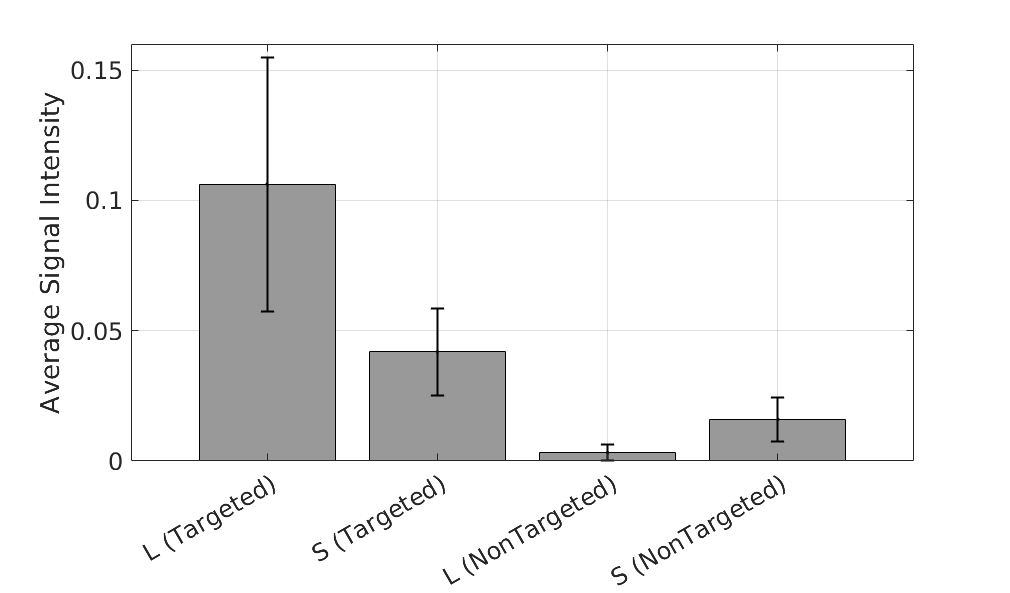}
    \end{subfigure}
    \caption{Average signal intensity of \textcolor{black}{$L$ and $S$ images} for targeted and non-targeted microbubbles.}
    \label{fig:invivo_signal_intensity_plots}
\end{figure}

Figure~\ref{fig:invivo_images_nonTargeted} shows the results of injecting nontargeted microbubbles in two different mice.
The first and second rows correspond to mouse 1, while the third and fourth rows correspond to mouse 2. For each mouse, the original contrast image, the DTE image, the estimated low-rank image, and the sparse image obtained from RPCA are demonstrated.
Because nontargeted microbubbles do not bind to molecular targets, they remain freely circulating in the bloodstream and do not accumulate in tumor regions. This results in little to no signal in the tumor area compared to targeted microbubbles as shown in both DTE and low-rank images. \textcolor{black}{Movies showing the time series contrast mode images and corresponding low-rank and sparse components for several in vivo mouse examples are included in the supplementary material.}

 bound and free microbubbles
 Figure~\ref{fig:invivo_signal_intensity_plots} shows the average signal intensity of \textcolor{black}{$L$ and $S$ images} within the tumor region across all mice. \textcolor{black}{The L and S images are consistent with the expected characteristics of bound and unbound microbubbles:} For the targeted microbubbles, the signal intensity in the \textcolor{black}{L image} is higher than in the \textcolor{black}{S image, as the former represents bound microbubbles while the latter represents free microbubbles. In contrast, for the non-targeted case, the opposite trend is observed,} as expected. \textcolor{black}{Furthermore,} the signal intensity of the targeted microbubbles is higher in both the \textcolor{black}{L and S images} compared to the non-targeted case.

\subsection{Execution Time}

There is a trade-off between the number of frames used in \textcolor{black}{the estimation of $L$ and $S$ matrices, the resulting accuracy, and the computational time}. \textcolor{black}{While a higher number of frames improves accuracy, it also increases the algorithm's execution time}. The average running time to decompose a data matrix consisting of 20 frames, each of size 338 by 260 \textcolor{black}{pixels}, into time series of low-rank and sparse matrix sequences of the same size is 0.2~s on a single NVIDIA GeForce RTX 3090 GPU. The average running times for 60 frames and 132 frames are 0.3~s and 0.4~s, respectively.

\section{Discussion}
\label{sec:discussion}

One of the bottlenecks for real-time and nondestructive UMI is separating the free and bound microbubbles, as detecting the latter is the main objective. In this work, we proposed using the RPCA method for this purpose and validated it using simulation, phantom, and in vivo data. Although the method has been demonstrated \textcolor{black}{in the context of molecular imaging of} breast cancer, its underlying principles are not limited to this specific application. Since the approach focuses on distinguishing free and bound microbubbles, it can potentially be adapted for \textcolor{black}{molecular imaging of other} organs, such as the liver, kidneys, or cardiovascular system.

\textcolor{black}{In the RPCA method, there is a trade-off between the number of frames used in reconstruction, accuracy, and run time. Using more frames yields more accurate results at the cost of longer running times.} For simulation data, the RMS error and Dice coefficient improve as the number of frames used in \textcolor{black}{the estimation of $L$ and $S$ matrices} increases. The RMS error of the $S$ matrix is lower than that of the $L$ matrix due to the smaller number of microbubbles and a higher proportion of zero values in the image. The Dice coefficient of $S$ is slightly better than that of  $L$ for the same reason. Increasing the number of frames decreases both the average and standard deviation (STD) of the RMS error for the $L$ matrix. However, for the $S$ matrix, only the STD decreases, while the average RMS error remains relatively constant. This is because ultrasound thermal noise can appear in the $S$ image, \textcolor{black}{limiting the lower bound on the} reconstruction \textcolor{black}{error}.  

\textcolor{black}{For the in vivo data, although using more frames improves accuracy, it also increases susceptibility to tissue or probe motion.}
\textcolor{black}{If the breathing} motion affects the entire background slowly and coherently over time, it is captured in the low-rank component ($L$). 
\textcolor{black}{In this work, for the reference results, we eliminated motion, such as respiratory motion, before applying the RPCA filter.} As future work, performing motion compensation before RPCA to align frames and reduce the impact of breathing motion, or using high-pass temporal filters before RPCA to remove slow-varying background motion, might be helpful and can be compared with the current results. 

The method relies on having a sufficient number of temporal frames to achieve good performance. Using too few frames (e.g., fewer than 5 frames) results in high RMSE and low Dice values, making detection less reliable. Furthermore, there is a trade-off between accuracy and computation: More frames improve separation accuracy but increase computational cost and memory usage, as the singular value thresholding step in RPCA is computationally expensive, especially when processing a large number of frames.

Based on the simulation results, using 20 frames to construct the data matrix is a practical choice, balancing Dice (0.75) and RMSE (0.1) while keeping computational time (0.2~s) \textcolor{black}{and motion} manageable. If computational cost is not \textcolor{black}{of} concern, increasing the number of frames to 30–40 could slightly improve \textcolor{black}{filtering of the image}. \textcolor{black}{The} simulation results showed that using a higher number of frames improved \textcolor{black}{separation of the stationary signals from the moving signals}.

The proposed method is non-invasive and can remove the moving microbubbles, but it cannot fully eliminate tissue leakage signal, as it appears in the contrast image background and decreases the Dice value \textcolor{black}{(Figure 6(c))}. DTE effectively cancels tissue leakage but \textcolor{black}{it cannot work in real-time and destroys the microbubbles. furthermore, reducing the number of frames in the averaging for DTE makes it more susceptible to moving bubbles.} 
\textcolor{black}{A limitation of the in vivo study is the absence of a ground-truth for the experiments. To address this, we used a large number of frames (132) \textcolor{black}{and removed those affected by motion before applying RPCA, which was then used as a reference. This choice was} based on simulation results indicating that a higher number of frames improves the accuracy of separating free and bound microbubbles. However, this reference does not completely eliminate tissue leakage signals, as previously mentioned, and therefore does not perfectly align with DTE. Despite this, it shows good agreement with DTE, and the differences between using 132 frames versus 20 or 60 frames were relatively small. This supports the use of 20 frames for RPCA filtering and imaging.}
A potential solution to these limitations would be to integrate the RPCA filter into our prior neural network-based approach~\cite{hyun2020nondestructive} to leverage the benefits of tissue leakage cancellation while also eliminating moving microbubbles.

The GPU-based implementation of RPCA enabled efficient decomposition of ultrasound molecular imaging data, significantly reducing computation time compared to a CPU-based approach. However, further optimization could improve performance and enable higher frame rates. While our implementation leverages CUDA acceleration through PyTorch, it primarily focuses on functionality rather than full optimization. Techniques such as mixed-precision computation, kernel fusion, and memory-efficient tensor operations could further accelerate processing~\cite{micikevicius2017mixed, chen2018tvm} \textcolor{black}{to achieve real-time performance.} Additionally, implementing a more hardware-aware approach, such as optimizing memory access patterns and utilizing tensor cores, could enhance performance~\cite{micikevicius2017mixed, guide2020cuda}.

\section{Conclusion}
\label{sec:conclusion}
We proposed a fast GPU-based method leveraging robust principal component analysis to differentiate bound microbubbles from free-floating ones. The method was validated using ten simulation datasets and phantom experiments. Furthermore, it was applied to data from ten transgenic mouse models of breast cancer development injected with B7-H3-targeted microbubbles, and two mice injected with non-targeted microbubbles. The results align well with the reference results, with a Dice coefficient higher than 0.95, and with those of the conventional differential targeted enhancement method, which shows a Dice value higher than 0.8. The average running time of the method on a data matrix of size 338 × 260 × 20 was 0.2~s on a single GPU.

\bibliographystyle{IEEEtran}
\bibliography{main}
\end{document}